\pdfoutput=1
\documentclass[%
 aip,
 amsmath,amssymb,
 reprint,%
]{revtex4-1}

\usepackage{graphicx}% Include figure files
\usepackage{dcolumn}% Align table columns on decimal point
\usepackage{bm}% bold math
\usepackage{comment}

\usepackage[utf8]{inputenc}
\usepackage[T1]{fontenc}
\usepackage{mathptmx}
\usepackage{etoolbox}
\usepackage{color}
\usepackage[normalem]{ulem}

\def\gs0{\gamma_\mathrm{S0}}

\def\ie{\textit{i.e.}, }

\def\eqref#1{(\ref{#1})}

\newcommand{\osaka}{Department of Mechanical Engineering, Osaka University, 2-1 Yamadaoka, Suita 565-0871, Japan}
\newcommand{\osakamu}{Department of Mechanical Engineering, 
Osaka Metropolitan University, 3-3-138 Sugimoto, Sumiyoshi, Osaka 558-8585, Japan}
\newcommand{\tuswater}{Water Frontier Research Center (WaTUS),
Research Institute for Science \& Technology,
Tokyo University of Science,
1-3 Kagurazaka, Shinjuku-ku, Tokyo, 162-8601, Japan}
\newcommand{\cnrs}{Univ Lyon, Univ Claude Bernard Lyon 1, CNRS, Institut Lumi\`ere Mati\`ere, F-69622, VILLEURBANNE, France}

\makeatletter
\def\@email#1#2{%
 \endgroup
 \patchcmd{\titleblock@produce}
  {\frontmatter@RRAPformat}
  {\frontmatter@RRAPformat{\produce@RRAP{*#1\href{mailto:#2}{#2}}}\frontmatter@RRAPformat}
  {}{}
}%
\makeatother
\begin{document}

%\preprint{AIP/123-QED}

\title[]{The receding contact line cools down during dynamic wetting}
% Force line breaks with \\
\author{Hiroki Kusudo}
\email{kusudo@tohoku.ac.jp}
\altaffiliation[Present Address: ]{Institute of Fluid Science, Tohoku University, 2-1-1 Katahira Aoba-ku, Sendai 980-8577, Japan}
\affiliation{\osaka}
%\affiliation{\tohoku}
%
\author{Takeshi Omori}
%\email{t.omori@omu.ac.jp}
\affiliation{\osakamu}
\author{Laurent Joly}
%\email{laurent.joly@univ-lyon1.fr}
\affiliation{\cnrs}
\author{Yasutaka Yamaguchi}
%\email{yamaguchi@mech.eng.osaka-u.ac.jp}
\affiliation{\osaka}
\affiliation{\tuswater}

\date{\today}% It is always \today, today,

\begin{abstract}
When a contact line (CL) ---where a liquid-vapor interface meets a substrate--- is put into motion, it is well known that the contact angle differs between advancing and receding CLs. 
Using non-equilibrium molecular dynamics simulations, 
we reveal another intriguing distinction between advancing and receding CLs: while temperature increases at an advancing CL ---as expected from viscous dissipation, we show that temperature can drop at a receding CL. 
Detailed quantitative analysis based on the macroscopic energy balance 
around the dynamic CL showed that the internal energy change of the fluid along the pathline
%, identical to streamline in steady state,
induced a remarkable temperature drop around the receding CL, in a manner similar to latent heat upon phase changes. 
This result provides new insights for modeling the dynamic CL, and the framework for heat transport analysis introduced here can be applied to a wide range of nanofluidic systems.
\end{abstract}

\maketitle

\section{Introduction}
Wettings are ubiquitous in our daily life, in nature and in various scientific and engineering fields. 
In particular, the behavior of the contact line (CL), where a liquid-vapor interface meets a solid surface, 
has long been a topic of interest
because it plays a key role in wetting properties.~\cite{deGennes2003,Ono1960,Rowlinson1982,Drelich2020} 
For {static} wetting without CL motion, a common measure of wettability at the macroscopic scale is the contact angle (CA), 
described by Young’s equation,~\cite{Young1805} 
which was first proposed in 1805 based on a 
balance 
between  
solid-liquid, solid-vapor and liquid-vapor interfacial tensions. 
These interfacial tensions originate from the microscopic molecular interaction forces, and 
recent molecular simulation studies have 
provided significant advance in understanding {static} wetting.~\cite{Nijmeijer1990, Tang1995, VanRemoortere1999,  Ingebrigtsen2007, Seveno2013, Leroy2009, Leroy2010, Leroy2015,  Ardham2015, Kanduc2017, Kanduc2017a, Jiang2017, Ravipati2018, %Surblys2019, Nishida2014, Yamaguchi2019, Bistafa2021, Kusudo2019, Imaizumi2020, Watanabe2022, Teshima2022, 
Smith2020, Heier2021, Bey2020, Weijs2011, Weijs2013, Marchand2012}
\par
The situation is more complex at the dynamic CL (DCL) ---appearing typically during droplet spreading or moving on a substrate, where the advancing and receding CAs are different.
To model the CA difference, numbers of theoretical, {computational} and experimental studies about the DCL have been carried out and have indicated that 
this dynamic effect is induced by the viscosity and friction 
in the vicinity of the DCL;~\cite{Hizumi2015,Omori2019,Thompson1989,Feng2018,Sefiane2008,Blake1969,Blake2002,Qian2003,Qian2006,deGennes2003,Thalakkator2020,Fernandez-Toledano2019,Fernandez-Toledano2021,Qian2003,Qian2006,Seveno2011,Chen2022,Smith2018,Lacis2020,deGennes2003,DeGennes1985,Voinov1977,Snoeijer2013,Cox1986,Huh1971,Dussan1979,Bonn2009,Snoeijer2013,Blake2019} 
however, 
the governing principle of the DCL motion still remains unclear, 
mainly due to the lack of detailed information on the nanoscale 
thermal and flow fields around the DCL, 
and it is considered to 
be one of the long-standing unsolved problems of fluid dynamics.
\par
In this article, 
we show a unique thermal phenomenon around the DCLs, 
cooling as well as heating at the DCLs. 
To elucidate 
its mechanism,
we analyze the heat flow field around the DCL using molecular dynamics (MD) simulations of a quasi-2D system with liquid–solid–vapor CLs, consisting of a {Lennard-Jones (LJ)} fluid between parallel solid walls moving in opposite directions 
as shown in the top panel of Fig.~\ref{fig:fig1}. 
To that aim, 
we have developed a heat transport analysis methodology applicable in multi-component MD systems.

\section{Methodology}
Prior to the analysis, 
we first need to construct a methodology to calculate heat flows based on the 
method of planes (MoP),~\cite{Todd1995, ToddDaivis1995, Todd_Daivis2017, Zhang2004, Smith2017,shi_perspective_2023, Heyes2011} 
which defines surface-averaged field values on a finite control plane 
so that obtained values satisfy the continuum conservation laws described by the Reynolds transport theorem for arbitrary control volume (CV) surrounded by finite control planes.~\cite{Kusudo2021} 
Specifically, in this article, we extend the formulation proposed 
for single-component fluid systems by \citet{Todd_Daivis2017} 
to the heat flow in multi-component systems with a solid wall. 
Energy conservation in the presence of an external force writes
\begin{align}
\label{eq:enecon}
\frac{\partial \rho e}{\partial t} = -\nabla \cdot  ( \rho e \bm{u} + \bm{J}_{\text{Q}}-\bm{\tau}\cdot \bm{u})+\rho \bm{F^{\text{ext}}}\cdot \bm{u},
\end{align}
where $\rho$, $\bm{u}$ and $e$ denote the density, velocity and specific total energy of the fluid ---defined by 
the sum of the specific internal energy and the specific convective kinetic energy $\frac{1}{2} |\bm{u}|^{2}$, 
whereas $\bm{J}_{\text{Q}}, \bm{\tau}$ and $\bm{F}^{\text{ext}}$ denote the heat flux, stress tensor and external force per unit mass, respectively.
For this energy conservation law, we treat the fluid--fluid intermolecular interaction force 
as the stress while we treat the fluid--solid one as the external force.~\cite{Schofield1982,Rowlinson1993} 
Equation~\eqref{eq:enecon} can be integrated for an arbitrary CV, and by applying Gauss' theorem to the advection and stress work terms on the right hand side (RHS), 
one obtains:
\begin{align}
    \int _{\text{CV}} \text{d} V 
   \frac{\partial \rho e}{\partial t} 
    =    
    &-
    \int _{\text{CV}} \text{d}V 
    \nabla \cdot \bm{J}_{\text{Q}}    
       +
    \int _{\text{CV}} \text{d}V 
    \rho \bm{F}^{\text{ext}} \cdot \bm{u}
    \nonumber \\
   &-
    \int _{\text{S}} \text{d} \bm{S} \cdot \left(
    \rho e\bm{u} 
    -
    \bm{\tau} \cdot \bm{u}
    \right),
    \label{eq:ene_con_int}
\end{align}
meaning that 
the fluid energy change in the CV in the {left hand side (LHS)} 
balances the heat production/absorption, the work of the external body force on the fluid in the CV, and 
the macroscopic energy advection and stress work through its surrounding surface in the RHS.
The divergence of the heat flux term, which corresponds to the heat production/absorption value in the CV, can be rewritten as
\begin{align}
    \int _{\text{CV}} \text{d}V \nabla \cdot
    \bm{J}_{\text{Q}}
    =
    &
    -
    \int _{\text{CV}} \text{d}V 
    \frac{\partial \rho e}{\partial t}
    +
    \int _{\text{CV}} \text{d}V 
    \rho \bm{F}^{\text{ext}} \cdot \bm{u}^{\text{VA}}
    \nonumber \\
    &
    -
    \int _{\text{S}} \text{d} \bm{S} \cdot 
    \rho e\bm{u} 
    +
    \int _{\text{S}} \text{d} \bm{S} \cdot 
    \bm{\tau} \cdot \bm{u},
    \label{eq:ene_flow_surface}
\end{align}
meaning that the heat flow from the CV is obtained 
by integrating the energy change in the CV and the energy advection and stress work on the surface of the CV 
---obtainable by the MoP, and by integrating the work by the external body force on fluid in the CV. 
In this article, we calculated the first term in the RHS by integrating the energy flux through the whole surrounding surface of the CV, 
see detail in the supplementary materials (SM).
%Appendix~\ref{sec:AppB}. 
Note that we adopted the volume-averaged fluid velocity $\bm{u}^{\text{VA}}$ in the second term of the RHS of Eq.~\eqref{eq:ene_flow_surface}
because we take its inner product with the body force $\rho \bm{F}^{\text{ext}}$ as the volume-averaged intermolecular force exerted on the fluid by the solid.

\section{System}

\begin{figure}
  \begin{center}
    \includegraphics[width=\linewidth]{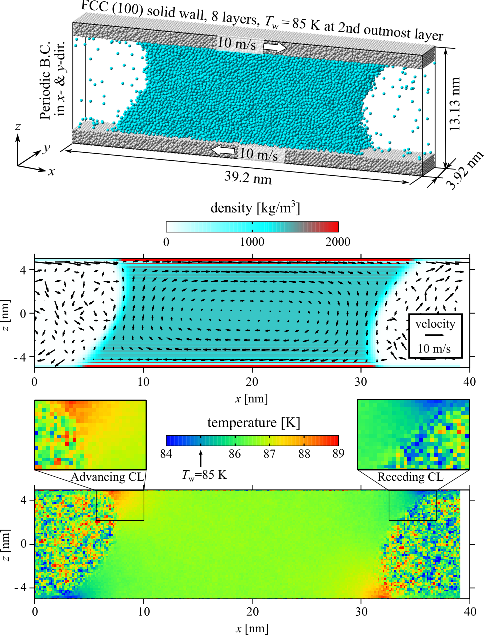}
  \end{center} 
  \caption{\label{fig:fig1}
Top: quasi-2D Couette-type flow system of a Lennard-Jones liquid confined between two solid walls. 
Middle: density and velocity distributions calculated by volume average; the macroscopic streaming velocity is denoted by black arrows. Bottom: temperature distribution of the fluid. (Partly from Kusudo, H., Omori, T., Yamaguchi, Y., J. Chem.
Phys., 155 (2021), 184103; licensed under CCBY)
} 
\end{figure}
The top panel of Fig.~\ref{fig:fig1} shows the MD simulation system of a quasi-2D Couette-type flow, where 
the basic setups are the same as in our previous study.~\cite{Kusudo2021} 
The fluid--fluid and fluid--solid interactions were modeled by the 12-6 LJ potential 
$\Phi^\mathrm{LJ}(r_{ij}) =  4\epsilon_{ij} \left[ \left(\frac{\sigma_{ij}}{r_{ij}}\right)^{12}-
\left(\frac{\sigma_{ij}}{r_{ij}}\right)^{6} \right]$, 
where $r_{ij}$ is the distance between the particles $i$ and $j$, while $\epsilon_{ij}$ and $\sigma_{ij}$ denotes the LJ energy and length parameters, respectively.
Quadratic functions were added to this LJ potential so that the potential and interaction force smoothly vanished at a cut-off distance of $r_\mathrm{c}=3.5 \sigma$.~\cite{Nishida2014} 
We used the following parameters for fluid--fluid (ff) and fluid--solid (fs) interactions:
$\sigma_{\text{ff}}=0.340$\,nm, $\epsilon_{\text{ff}}=1.67\times10^{-21}$\,J, 
$\sigma_{\text{fs}}=0.345$\,nm, $\epsilon_{\text{fs}}=0.646\times10^{-21}$\,J.
The atomic masses of fluid and solid particles were $m_{\text{f}}=39.95$\,u and $m_{\text{s}}=195.1$\,u, respectively.
Finally, the equations of motion were integrated using the velocity-Verlet algorithm, with a time step $\Delta t$ of 5\,fs.

Periodic boundary conditions were set in the $x$- and $y$-directions, and 20,000 LJ particles were 
confined between two parallel solid walls (dimension of $x \times y = 39.2\times 3.92$\,nm$^{2}$) at a distance of $\sim 10.4$\,nm, 
so that the LJ fluid {formed} two quasi-2D menisci with CLs on the walls upon the 
preliminary equilibration at a control temperature $T_{\mathrm{w}}=85$\,K without shear. 
The static CA on both top and bottom walls was $\sim 57$\,deg.
 After the equilibration, further relaxation runs to achieve a steady shear flow with asymmetric menisci were carried out for 10\,ns by moving the particles in the outmost layers of both walls with opposite velocities of $\pm$10\,m/s in the $x$-direction. 
\par
After the relaxation run, 
the main calculation was conducted for an average time of 400\,ns. 
We calculated the external body force and volume averaged velocity in the RHS of Eq.~\eqref{eq:ene_flow_surface}
using cuboid bins of size 
$\Delta x \times \Delta y \times \Delta z=0.150 \times 3.92 \times 0.149  \,\text{nm}^3$, 
while we calculated the energy flux, 
velocity, stress and the specific energy in the RHS of Eq.~\eqref{eq:ene_flow_surface} using the MoP with the faces of each local bin. 
Regarding the calculation of the energy flux and the specific energy, see details in SM.% Appendix~\ref{sec:AppA} and ~\ref{sec:AppB}.

\section{Results and Discussion}
\begin{figure*}
  \begin{center}
    \includegraphics[width=.9\linewidth]{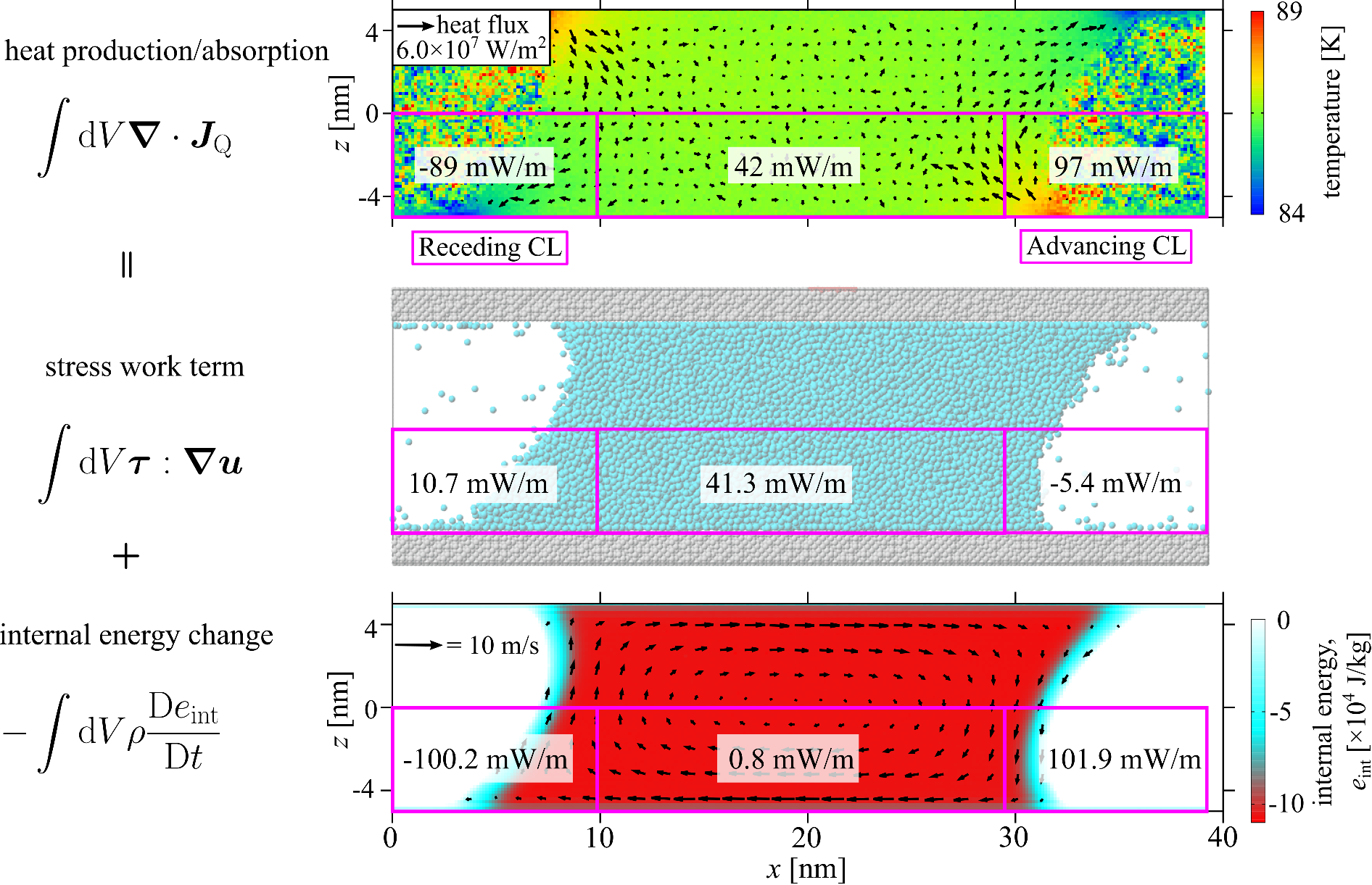}
  \end{center} 
  \caption{\label{fig:fig2}
  Volume integral of (top) the heat flux divergence, (middle) the inner product of stress tensor and tensor derivative of velocity vector and (bottom) the Lagrange derivative of internal energy on three CVs; left: surrounding the receding CL, right: surrounding the advancing CL, middle: between the left and right CVs.
  The top panel also shows the distributions of temperature and heat flux field.
  The bottom panel also shows the internal energy distribution and the velocity field.
}
\end{figure*}
The middle panel of Fig.~\ref{fig:fig1} shows the density distribution and velocity field obtained by the volume average. 
Due to the shear applied by the wall, a caterpillar-like flow was induced, and DCLs, \ie advancing and receding CLs with different CAs, appeared. 
In addition to the CA difference, we showed the stress inhomogeneity in the bulk liquid induced by this flow in our previous study.~\cite{Kusudo2021} 
In the present study, we report a distinct thermal difference in the DCLs as shown in the temperature distribution in the bottom panel of Fig.~\ref{fig:fig1}: 
temperature rises around the advancing CLs (bottom right and top left), and temperature drops around the receding CLs (bottom left and top right). 
Quantitatively, in the bulk liquid away from the interfaces, the temperature is around 86.5\,K, which is slightly higher than the control temperature of the wall due to viscous dissipation, 
whereas that around the advancing CLs is about 2\,K higher and that around the receding CLs is about 2\,K lower than the bulk, as shown in Fig.~\ref{fig:fig1}. 
The cooling at the receding CLs is especially intriguing, 
because viscous dissipation can only induce temperature rise through heat production.

To elucidate the mechanism of the heat production/absorption around the DCLs, 
we conducted a heat flux analysis. 
The top panel of Fig.~\ref{fig:fig2} shows the heat flux field superimposed 
on the temperature distribution.
Note that the heat flux field was depicted only for the fluid sufficiently away from the wall,  
where the effect of potential field from the wall on the fluid was negligibly small; the MoP methodology of the heat flux calculation is shown in SM. %Appendix~\ref{sec:AppC}.
Heat flow from the high temperature area to the low temperature area can be observed, 
meaning that the heat produced around an advancing CL induces temperature rise there, and flows to the cold neighboring receding CL due to the heat absorption. 
To quantitatively evaluate the heat production/absorption around the DCLs and in the liquid area sufficiently away from the CLs, 
we set three control volumes (CVs) as shown with magenta rectangles in Fig.~\ref{fig:fig2}; 
a CV surrounding the receding CL, a CV surrounding the advancing CL, and a CV between them. 
We integrated the divergence of the heat flux in each CV 
and the corresponding values are 
shown inside the CVs with a unit of mW/m, calculated as the heat production/absorption rate divided by the system depth, in the top panel of Fig.~\ref{fig:fig2}. 
According to Fig.~\ref{fig:fig2}, 
heat is produced and absorbed at the CVs surrounding the advancing ($97$\,mW/m) and receding ($-89$\,mW/m) CLs, respectively, 
and the absolute values are approximately twice as large as the heat production $42$ mW/m in the middle CV even though its volume is twice as large as the others. 
From this, we see that viscous dissipation is not the main cause of heat production/absorption at the DCLs. 
\par
To elucidate this, 
one can rewrite the energy conservation in Eq.~\eqref{eq:enecon} as follows (see details in SM): 
\begin{equation}
\label{eq:ene_cons2}
    \bm{\nabla}\cdot \bm{J}_{\mathrm{Q}}
    =
    \bm{\tau}:\bm{\nabla}\bm{u}
    -\rho \frac{\mathrm{D} e_{\mathrm{int}}}{\mathrm{D}t},
\end{equation}
where $e_{\mathrm{int}}$ denotes the specific internal energy defined by
\begin{align}
    e_{\mathrm{int}}
    &=
    e-\frac{1}{2}|\bm{u}|^{2} , 
    \label{eq:eint}
\end{align}
and $\frac{\mathrm{D}}{\mathrm{D} t}$ and $:$ denote {the} Lagrangian derivative and the inner product of a second order tensor, respectively. 
%\add{
Note that the 
specific internal energy includes not only the fluid--fluid interaction potential but also the fluid--solid one. 
Equation~\eqref{eq:ene_cons2} indicates that the heat production/absorption arises from two mechanisms: 
1. the inner product of stress tensor and velocity gradient, which corresponds to the viscous dissipation in bulk; 
2. the internal energy change along the pathline (identical to the streamline in a steady flow). 
The work done by the solid--fluid interaction force seems to be absent from Eq.~\eqref{eq:ene_cons2} because it does not contribute to the internal energy change but to the convective kinetic energy change, see details in SM. 
%Therefore, it can be explicit with the convective kinetic energy in the first term on the RHS. % (see details in Ref.~\citenum{Evans2008}).
%Also note that the work done by the solid on the fluid is included in the first term of the RHS of Eq.~\eqref{eq:ene_cons2}, where the solid--fluid interaction force is treated as the external force, \ie is not included in stress.
We rewrite the first term on the RHS of Eq.~\eqref{eq:ene_cons2} as
\begin{align}
    \bm{\tau}:\bm{\nabla}\bm{u}
    &=
    \bm{\nabla} \cdot (\bm{\tau} \cdot \bm{u} ) 
    -\bm{u} \cdot (\nabla \cdot \bm{\tau}) 
    \nonumber \\
    &=
    \bm{\nabla} \cdot (\bm{\tau} \cdot \bm{u} ) 
    -
    \frac{\partial }{\partial t} \frac{1}{2}\rho |\bm{u}|^{2}    
    -
    \bm{\nabla} \cdot \frac{1}{2}\rho |\bm{u}|^{2} \bm{u} \nonumber \\
    &
    \;\;\;\;\;\;\;\;\;\;\;\;\;\;\;\;\;\;\;\;\;\;\;\;\;\;\;\;\;\;\;\;\;\;\;\;\;\;\;\;
    \;\;\;\;\,
    +\rho \bm{F}^{\mathrm{ext}} \cdot \bm{u} \nonumber \\
    &=
    \bm{\nabla} \cdot (\bm{\tau} \cdot \bm{u} )
       -
    \bm{\nabla} \cdot \frac{1}{2}\rho |\bm{u}|^{2} \bm{u}
     +\rho \bm{F}^{\mathrm{ext}} \cdot \bm{u},
     \label{eq:6}
\end{align}
where the momentum conservation with external force is applied for the second equality and
macroscopically steady-state system is assumed for the third equality. 
The rightmost-hand side (HS) can be directly integrated for the CV with MoP by applying Gauss' theorem 
while the numerical differentiation is essential to integrate the LHS. 
The middle panel of Fig.~\ref{fig:fig2} shows 
the integral of this stress term by using Eq.~\eqref{eq:6}, 
and this indicates that this stress work is the main cause of the heat production in the middle CV without CL, 
but is not remarkably large in the CV surrounding the DCLs. 
Note that the stress term contains not only the viscous dissipation but also the work by the pressure or interfacial tensions
so that it is not always positive, specifically around the advancing CL.
The other 
factor of the heat production/absorption is 
the internal energy change $\rho \frac{\mathrm{D} e_{\mathrm{int}}}{\mathrm{D}t}$ in the RHS of Eq.~\eqref{eq:ene_cons2}, and we rewrite it as
\begin{align}
    -\rho \frac{\mathrm{D} e_{\mathrm{int}}}{\mathrm{D}t}
    &=
    -
    \frac{\partial \rho e_{\mathrm{int}} }{\partial t}
    -
    \bm{\nabla} \cdot \rho e_{\mathrm{int}} \bm{u}
    \nonumber \\
    &=
    -
    \frac{\partial \rho e }{\partial t}
    +
    \frac{\partial }{\partial t} \frac{1}{2}\rho |\bm{u}|^{2}
    -
    \bm{\nabla} \cdot \rho e \bm{u}
    +
    \bm{\nabla} \cdot \frac{1}{2}\rho |\bm{u}|^{2} \bm{u}
    \nonumber \\
    & =
        -
    \frac{\partial \rho e }{\partial t}
        -
    \bm{\nabla} \cdot \rho e \bm{u}
    +
    \bm{\nabla} \cdot \frac{1}{2}\rho |\bm{u}|^{2} \bm{u},
    \label{eq:7}
\end{align}
where the mass conservation is applied for the first equality, 
Eq.~\eqref{eq:eint} is applied for the second equality, and the convective kinetic energy is assumed to be constant over time.
Note that the steady state is not assumed for the first term in the rightmost-HS 
because it depends on the microscopic configuration difference between the start and end time of the sampling interval, and it is not negligibly small especially around the CLs, which are microscopically fluctuating.~\cite{Kusudo2021} 
Also for this Eq.~\eqref{eq:7}, the rightmost-HS can be directly integrated for the CV with the MoP by applying Gauss' theorem, 
and the integral of the first term in the rightmost-HS is obtained with the energy flux through the whole surrounding surface
of the CV, see detail in SM. %Appendix~\ref{sec:AppB}. 
The integral values of this internal energy change 
for the CVs are shown in the bottom panel of Fig.~\ref{fig:fig2}.
The large absolute values of $-100.2$ and $101.9$~mW/m in the two CVs indicate that this term is the main cause of the heat production/absorption around the DCLs, 
which is small in the middle CV without CL.

We also show the internal energy distribution and the velocity field as the background of bottom panel of Fig.~\ref{fig:fig2}, 
and one can observe that the internal energy changes along the streamline specifically near the DCLs, where the heat is produced/absorbed.
At the advancing CL,
heat is produced when the fluid flows from the solid-vapor and liquid-vapor regions to the solid-liquid region, 
whereas at the receding CL heat is absorbed when the fluid flows from the solid-liquid region to the solid-vapor and liquid-vapor regions.
During these processes, the internal energy of the fluid changes due to the surrounding density change as well as due to the potential field induced by the solid surface, 
and it leads to the cooling and heating at the DCLs.
This phenomenon is analogous to latent heat, which induces the heat production/absorption upon the phase change. 
\par
\begin{figure}
  \begin{center}
    \includegraphics[width=.9\linewidth]{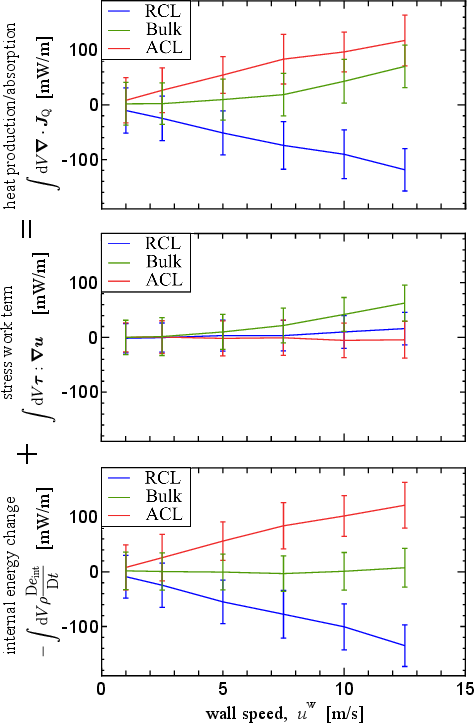}
  \end{center} 
  \caption{\label{fig:fig3}
     Volume integral values of (top) the heat flux divergence, (middle) the inner product of stress tensor and derivative of velocity vector and (bottom) the Lagrangian derivative of internal energy in Eq.~\eqref{eq:7} on the three CVs for various wall speed $u^{\text{w}}$. The locations of the CVs are the same as in Fig.~\ref{fig:fig2}. 
}
\end{figure}

Therefore, it is expected that this cooling and heating effect should be increased with the flow rate around the DCLs, \ie the faster wall speed.
Here, 
we additionally conducted the heat analysis for the CVs with various wall speeds: $u^{\text{w}}=$ 1.0, 2.5, 5.0, 7.5, and 12.5 m/s (the density and velocity fields and temperature distribution with each condition are shown in SM). % Appendix~\ref{sec:AppE}.
Top, middle and bottom panels of Fig.~\ref{fig:fig3} show the volume integral values of 
the heat flux divergence, stress work term in Eq.~\eqref{eq:6} and internal energy change along the streamline in Eq.~\eqref{eq:7}, respectively.
Blue and red ones denote the values of CV including the receding CL (RCL) and the advancing CL (ACL), and green one denotes the CV between them.
Note that the CV arrangement for all wall speeds is same as that in  Fig.~\ref{fig:fig2}.
Similar to Fig.~\ref{fig:fig2}, 
the internal energy change is the main part of heat production/absorption in CVs including DCLs 
while the stress work term is dominant in the middle CV (referred to as "Bulk" in Fig.~\ref{fig:fig3}).
The internal energy change appears to be proportional to the wall speed, implying that 
the spatial distributions of the density and the specific energy do not largely change due to the wall speed. 
On the other hand, the stress work term appears to be proportional to the square of the wall velocity in the middle CV 
since the shear stress, \ie viscous stress, is proportional to the shear rate in the bulk where the shear rate can be roughly proportional to the wall velocity. 
Also in the CVs including the DCLs, the work done by the solid--fluid interaction force in Eq.~\eqref{eq:6} should largely depend on the wall speed 
because that frictional force is supposed to be proportional to the slip velocity. 
Under the present wall velocities where the steady-state caterpillar-like flows with DCLs are achieved, the internal energy change is always dominant over the stress work term 
in the CVs including DCLs: and thus the temperature rise/drop near the DCLs should always exist. %, see the details in SM.
In addition, we observed this cooling/heating phenomena at the DCLs induced by the same mechanism also on less wettable walls as shown in SM. %Appendix~\ref{sec:AppD}.
%
%
%
%
%
%
%
%
%
%Appendix~\ref{sec:AppF}.
Note that this quasi-latent heat around the DCLs is not a dissipation energy, meaning that it cannot be included in the dissipation terms of existing macroscale DCL models.~\cite{DeGennes1985,Voinov1977,deGennes2003} 
However, it indeed induces temperature changes in the vicinity of the DCLs, which should be included in the DCL models.

\section{Conclusion}
In this article, we have presented a heat transport analysis methodology applicable in multi-component MD systems, which we have used to investigate the heat transport features of the DCL. 
The heat analysis revealed that heat is not only produced but also absorbed around DCLs, 
mainly due to the quasi-latent heat induced by the internal energy change of fluid along the pathline,
when the fluid moves among the interfaces, which is accompanied by a change in fluid--fluid and fluid--solid interaction energy.
In addition, this latent heat is not a dissipation energy, thus
almost the same heat is absorbed and produced at receding and advancing CLs, respectively,  
while heat is only produced in bulk liquid due to viscous dissipation. 
Overall, these results provide new insights into the molecular mechanisms controlling the dynamics of the CL. 
Moreover, the framework for analyzing the heat transport at the molecular scale should be useful for investigating various nanoscale systems such as the flow in carbon nanotubes or in nanoporous media.~\cite{Ebrahimi2018,Goldsmith2010,Toghraie2019,Rahmatipour2017,Noorian2014,Li2010,Thomas2010,Thompson2003}

\section*{SUPPLEMENTARY MATERIAL}
The supplementary material contains the calculation methods of the energy density, energy flux and heat flux by the method of plane and the derivation of the Lagrangian derivative of the internal energy in Eq.~\eqref{eq:ene_cons2}.
We also show therein the density, velocity and temperature distributions with various wall speeds on the lyophilic walls corresponding to Fig.~\ref{fig:fig3}, and
the density and temperature distributions around the DCL on the lyophobic walls.

\begin{acknowledgments}
HK, TO and YY are supported by JSPS KAKENHI Grant Nos. JP23KJ0090, JP23H01346 and JP22H01400, respectively. YY is also supported by JST CREST Grant No. JPMJCR18I1, Japan.
Numerical simulations were performed on the Supercomputer system "AFI-NITY" at the Advanced Fluid Information Research Center, Institute of Fluid Science, Tohoku University.
\end{acknowledgments}

\section*{Conflict of Interest Statement }
The authors have no conflicts to disclose.

\section*{Data Availability Statement}

The data that support the findings of this study are available from the corresponding author upon reasonable request.

%\nocite{*}
\bibliography{reference}% Produces the bibliography via BibTeX.

\end{document}

%\preprint{AIP/123-QED}

\title[]{Supplementary Materials -- The receding contact line cools down during dynamic wetting}
% Force line breaks with \\
\author{Hiroki Kusudo}
\email{kusudo@tohoku.ac.jp}
\altaffiliation[Present Address: ]{Institute of Fluid Science, Tohoku University, 2-1-1 Katahira Aoba-ku, Sendai 980-8577, Japan}
\affiliation{\osaka}
%\affiliation{\tohoku}
%
\author{Takeshi Omori}
%\email{t.omori@omu.ac.jp}
\affiliation{\osakamu}
\author{Laurent Joly}
%\email{laurent.joly@univ-lyon1.fr}
\affiliation{\cnrs}
\author{Yasutaka Yamaguchi}
%\email{yamaguchi@mech.eng.osaka-u.ac.jp}
\affiliation{\osaka}
\affiliation{\tuswater}

\date{\today}% It is always \today, today,

\maketitle

\section{Energy density calculation by the method of plane
\label{sec:AppA}}
The energy density $\rho e(S_{k},t)$ in the right hand side of Eq.(3)
can be obtained {by applying the method of plane (MoP) formulation 
shown in our previous study \cite{Kusudo2021}} as follows: 
\begin{equation}
\label{eq:enedens_ff}
    \rho e (S_{k},t) =   
    \lim_{\delta t \to 0}
    \frac{1}{S_{k}\delta t}\left<
    \sum_{i\in\mathrm{fluid},\delta t}^{\text{crossing } S_{k}}
    \frac{1}{|v_{k}^{i}|}
    e^{i}_{\text{f}}
    \right>,
\end{equation}
where $e^{i}_{\text{f}}$ and $v_{k}^{i}$ denote the molecular energy defined by Eq.~\eqref{eq:mol_ene_ff} 
and the $k$-component of the velocity vector $\bm{v}^{i}$ of fluid particle $i$, respectively.
The angular bracket denotes the ensemble average, 
and the summation $\sum_{i\in\mathrm{fluid},\delta t}^{\text{crossing } S_{k}}$ 
is taken for every fluid particle $i$ passing through $S_{k}$ within a time interval of $\delta t$, 
which is replaced by the time increment for the numerical integration. 

\section{The energy flux calculation by the method of plane
\label{sec:AppB}}
We calculated the first term in the RHS of Eq.(3) by integrating the energy flux through the whole surrounding surface of the CV, which is essential for the depiction of the heat flux,
while the integrated value is technically same as the total molecular energy change in the CV during the simulation.
The energy flux in the MoP formulation \cite{Todd_Daivis2017, Ohara1999} writes
\begin{align}
J_{\text{T}k}&
\equiv \lim_{\delta t \to 0}
    \frac{1}{S_{k}\delta t}
    \left<
    \sum_{i\in\mathrm{fluid}, 
    \delta t}^{\text{crossing} S_{k}}
    e_{\text{f}}^{i} \frac{v_{k}^{i}}{\left| v_{k}^{i}\right|} 
    \right>+
    \frac{1}{S_{k}}
    \left<
    \sum_{i \in \mathrm{fluid}, j \neq i }^{\text{across\ } S_{k}} 
    q^{ij} \frac{r^{ij}_{k}}{|r^{ij}_{k}|}
    \right>,
    \label{eq:MoP_ene_fluidfluid}
\end{align}
where $e_{\text{f}}^{i}$ denotes the fluid molecular energy defined by
\begin{align}
    e_{\text{f}}^{i}
    =
    \frac{1}{2}m|\bm{v}^{i}|^{2} 
    +
    \sum_{i \in \text{fluid}, j \neq i} 
    \frac{1}{2}\Phi_{ij}, 
    \label{eq:mol_ene_ff} 
\end{align}
and $r^{ij}_{k}$ denotes the relative position of $j$-th molecule from $i$-th.
In addition, $q^{ij}$ is the intermolecular energy transport from $i$-th to $j$-th defined by 
\begin{align}
q^{ij}= \frac{1}{2} \bm{F}^{ji} \cdot \left(\bm{v}^{j} +\bm{v}^{i}  \right),
\label{eq:mol_ene_trans}
\end{align}
where $\bm{F}^{ji}$ denotes the intermolecular force exerted on $j$-th from $i$-th molecule.
This is based on the mechanism that molecular kinetic energy changes due to the work done by the intermolecular force.
In the velocity-Verlet algorithm, the change of molecular energy at $t+\Delta t$ from time $t$ writes
%\begin{widetext}
\begin{align}
    \frac{1}{2}m|\bm{v}^{i}(t+\Delta t)|^{2}
    -
    \frac{1}{2}m|\bm{v}^{i}(t)|^{2}
    &=
    \frac{1}{2}m \left|
    \bm{v}^{i}(t)+
  {\Delta}t \cdot \frac{\bm{F}_i(t)
  +\bm{F}_i(t+{\Delta}t)}{2m} \right|^{2}
  -
    \frac{1}{2}m|\bm{v}^{i}(t) |^{2} \nonumber \\
    &=
    \left[
    \frac{\bm{F}_i(t)
  +\bm{F}_i(t+{\Delta}t)}{2}
    \right] 
    \cdot
    \left[ 
    \bm{v}^{i}(t)+  
    \frac{\Delta t}{2}
    \frac{\bm{F}_i(t)  +
    \bm{F}_i(t+{\Delta}t)}{2m} 
    \right]
    \Delta t,
\end{align}
%\end{widetext}
where $\bm{F}_{i}(t)$ and $\bm{F}_{i}(t+\Delta t)$ denote the total intermolecular force exerted on molecule $i$ at time $t$ and $t+\Delta t$, respectively.
We can interpret that the time derivative of kinetic energy at $t+\frac{\Delta t}{2}$
equals the inner product of the force and velocity, which are defined by
\begin{align}
    \bm{F}_{i} \left( t+\frac{\Delta t}{2} \right)
    &=
    \frac{\bm{F}_i(t)
  +\bm{F}_i(t+{\Delta}t)}{2}
\end{align}
and
\begin{align}
  \bm{v}^{i} \left(t+\frac{\Delta t}{2} \right)
    &=
    \bm{v}^{i}(t)+  
    \frac{\Delta t}{2}
    \frac{\bm{F}_i(t)  +
    \bm{F}_i(t+{\Delta}t)}{2m}, 
\end{align}
respectively.
To satisfy the relationship of molecular energy transport 
in the energy flux calculation in Eq.~\eqref{eq:MoP_ene_fluidfluid}, 
we rewrite Eq.~\eqref{eq:mol_ene_trans} as 
\begin{align}
q^{ij}= \frac{1}{2} \bm{F}^{ji}\left(t+\frac{\Delta t}{2}\right)  
\cdot 
\left[
\bm{v}^{j} \left(t+\frac{\Delta t}{2} \right)
+\bm{v}^{i} \left(t+\frac{\Delta t}{2} \right)
\right],
\end{align}
where 
we adopted 
for the intermolecular force 
\begin{align}
    \bm{F}^{ji}\left(t+\frac{\Delta t}{2}\right)&=
    \frac{\bm{F}^{ji}(t)
  +\bm{F}^{ji}(t+{\varDelta}t)}{2}.
\end{align}
This assignment of the force and velocity guarantees
that the energy change in the control volume (CV) equals to the integral of the energy flux for the whole surface of the CV calculated by Eq.~\eqref{eq:MoP_ene_fluidfluid}.

\section{Heat flux calculation away from the wall
\label{sec:AppC}}
The heat flux field in the top panel of Fig.~2 was obtained by using the MoP.
The energy conservation without the external force writes
\begin{align}
\label{eq:enecon_f}
\frac{\partial \rho e}{\partial t} = -\nabla \cdot  ( \rho e \bm{u} + \bm{J}_{\text{Q}}-\bm{\tau}\cdot \bm{u}),
\end{align}
and in this article it holds for the fluid sufficiently away from the wall, where the effect of the potential field from the wall on the fluid was negligibly small.
The energy change in the left hand side (LHS) in the control volume (CV) balances the energy flux through the surrounding surface of the CV.
Therefore, Eq.~\eqref{eq:enecon_f} writes
\begin{align}
\label{eq:enecon_f2}
 -\nabla \cdot \bm{J}_{\text{T}}= -\nabla \cdot  ( \rho e \bm{u} + \bm{J}_{\text{Q}}-\bm{\tau}\cdot \bm{u}),
\end{align}
where $\bm{J}_{\text{T}}$ denotes the energy flux.
By assuming that the energy flux in the LHS is comprised of the macroscopic advection term, heat flux and stress work terms in the RHS,
the heat flux through the bin face $S_{k}$
with its normal vector pointing to the $k$-th Cartesian direction
$J_{\text{Q}}^{\text{ff}}(S_{k},t)$ 
can be 
calculated from the other three terms in Eq.~\eqref{eq:enecon_f2} obtainable via the Method-of-Plane as
\begin{align}
\label{eq:heatflux_ff_MoP}
J_{\text{Q}}^{\text{ff}}(S_{k},t)=
J_{\text{T}}^{\text{ff}}(S_{k},t)
-
\rho e^{\text{ff}}(S_{k},t) u_{k} (S_{k},t) 
+ 
\tau_{kl}(S_{k},t) u_{l}(S_{k},t),
\end{align}
where $J_{\text{T}}^{\text{ff}}(S_{k},t)$, $\rho e^{\text{ff}}(S_{k},t)$, $ u_{k} (S_{k},t)$ 
and $\tau_{kl}(S_{k},t)$ denote the energy flux~\cite{Ohara1999,Torii2008,Todd_Daivis2017,ToddDaivis1995}
through the bin face $S_{k}$, 
the energy density and the averaged velocity~\cite{Kusudo2021} on the bin face $S_{k}$
and the stress in the $l$-direction exerted on the bin face $S_{k}$, respectively.
Note that 
the third term on the RHS of Eq.~\eqref{eq:heatflux_ff_MoP} denotes 
the inner product of the stress and the velocity vector on the bin face $S_{k}$, \ie the Einstein notation is used 
for the dummy index $l$. 

\section{Derivation of the Lagrangian derivative of the internal energy
\label{sec:AppF}}
In the main text, we analyzed the heat transport around the DCLs based on the 
the energy conservation in the Lagrangian form of the internal energy in Eq.~(4). 
Here, we show the derivation from the equations of mass and momentum conservation and Eq.~(5) as the definition of the internal energy $e_{\text{int}}$ (see also Ref.~\citenum{Evans2008}). 
The mass conservation writes
\begin{equation}
    \frac{\ptl \rho }{\ptl t}+\nabla \cdot \rho \bm{u} =0,
        \label{eq:mass}
\end{equation}
and the momentum conservation with the presence of the external force $\rho F^{\text{ext}}$ writes
\begin{equation}
    \frac{\ptl \rho \bm{u} }{\ptl t}+\nabla \cdot \rho \bm{u} \bm{u} = \nabla \cdot \tau + \rho F^{\text{ext}}.
    \label{eq:momentum}
\end{equation}
By introducing the Lagrangian derivative
\begin{equation}
\frac{D}{Dt} \equiv \frac{\ptl}{\ptl t} + \bm{u}\cdot \nabla,
\end{equation}
and by inserting mass conservation in Eq.~\eqref{eq:mass}, 
the momentum conservation 
in Eq.~\eqref{eq:momentum} writes 
\begin{equation}
    \rho \frac{D \bm{u} }{D t} = \nabla \cdot \tau + \rho F^{\text{ext}}.
    \label{eq:momentum2}
\end{equation}
By taking the inner product of Eq.~\eqref{eq:momentum2} with $\bm{u}$, 
the Lagrangian derivative of the convective kinetic energy $1/2 |\bm{u}|^2$ is derived as follows:
\begin{align}
    \bm{u} \cdot \rho \frac{D \bm{u} }{D t} &= \bm{u} \cdot  \nabla \cdot \tau + \bm{u} \cdot \rho F^{\text{ext}} \nonumber \\
    \rho \frac{D }{D t} \left(\frac{1}{2} |\bm{u}|^2\right)  &= \nabla \cdot \left( \tau \cdot \bm{u} \right) - \tau : \nabla \bm{u}+  \rho F^{\text{ext}} \cdot \bm{u}, 
    \label{eq:kineticene}
\end{align}
where : denotes the inner product of a second order tensor. This equation means that 
the inner product of stress tensor with the velocity gradient 
does not contribute to the change of kinetic energy because it corresponds to the viscous dissipation in the bulk, whereas the work done by the external force contributes to the change of kinetic energy. 

By transposing $\nabla \cdot \rho e \bm{u}$,  in Eq.~(1) as well, the Lagrangian form of the energy conservation writes 
\begin{align}
\rho \frac{D e }{D t} = -\nabla \cdot ( \bm{J}_{\text{Q}}-\bm{\tau}\cdot \bm{u})+\rho \bm{F^{\text{ext}}}\cdot \bm{u}, 
\label{eq:ene2}
\end{align}
meaning that the fluid energy change is induced by the heat flux divergence, the stress work and the work done by the external force. 
By inserting Eq.~(5), it follows 
\begin{align}
\rho \frac{D e_\text{int} }{D t} + 
\rho \frac{D }{D t} \left(\frac{1}{2}|\bm{u}|^2 \right)
= -\nabla \cdot ( \bm{J}_{\text{Q}}-\bm{\tau}\cdot \bm{u})+\rho \bm{F^{\text{ext}}}\cdot \bm{u}. 
\end{align}
By further inserting Eq.~\eqref{eq:kineticene}, 
the Lagrangian of the internal energy in Eq.~(4) is derived:
\begin{align}
\rho \frac{D e_\text{int} }{D t}  
= -\nabla \cdot \bm{J}_{\text{Q}} + \bm{\tau} :\nabla \bm{u}.
\end{align}
From this equation, it is shown that the work done by the external force is implicit, \textit{i.e.}, it does not contribute to the change of internal energy but to the change of convective kinetic energy. 
\clearpage
\section{Density, velocity and temperature distributions with various wall speed on the lyophilic wall
\label{sec:AppE}}
\begin{figure}[b]
  \begin{center}
    \includegraphics[width=\linewidth]{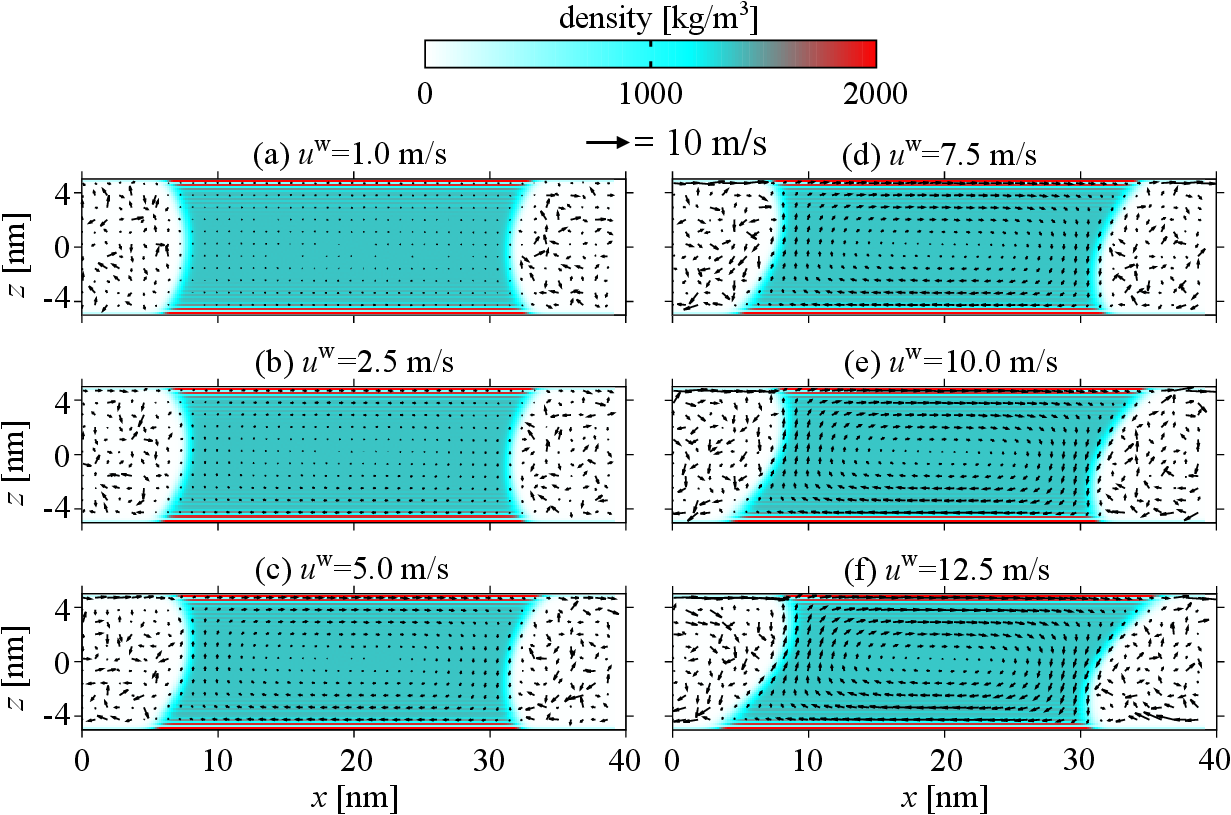}
  \end{center} 
  \caption{\label{fig:figs3}
  Density and velocity fields with various wall speeds: $u^{\text{w}}=$ 1.0, 2.5, 5.0, 7.5, 10.0, and 12.5 m/s on the lyophilic wall.
}
\end{figure}

\begin{figure}
  \begin{center}
    \includegraphics[width=\linewidth]{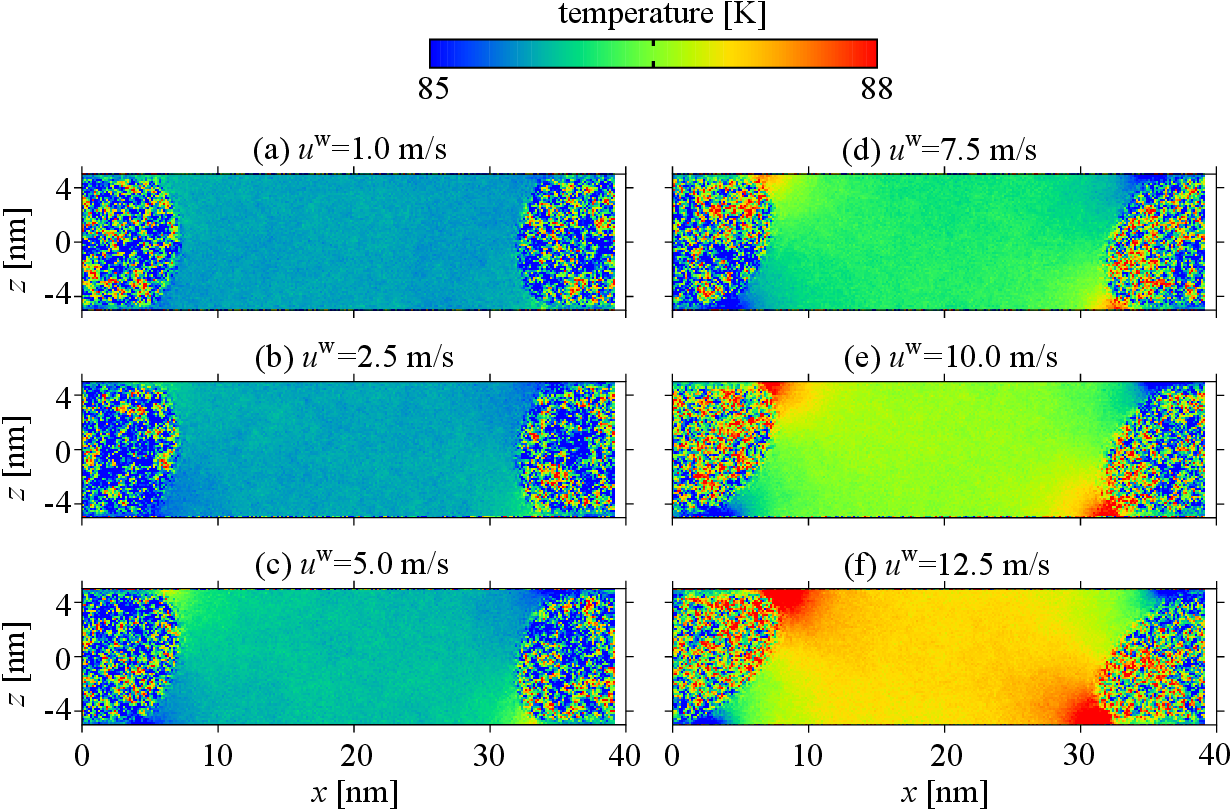}
  \end{center} 
  \caption{\label{fig:figs4}
  Temperature distribution with various wall speeds: $u^{\text{w}}=$ 1.0, 2.5, 5.0, 7.5, 10.0, and 12.5 m/s on the lyophilic wall. 
%  The magenta rectangles are the regions for obtaining the local DCL temperatures in Fig.~\ref{fig:figs2} in Appendix~\ref{sec:AppF}.
}
\end{figure}

We conducted the heat flow analysis of the DCLs with various wall speeds: $u^{\text{w}}=$ 1.0, 2.5, 5.0, 7.5, 10.0, and 12.5 m/s on the lyophilic wall.
The density and velocity fields and temperature distribution with each condition are shown in Fig.~\ref{fig:figs3}. 
With the increase of the wall speed, the speed of the caterpillar-like flow became larger and the meniscus shape got more skewed,
resulting in 
a larger difference in the CAs between the advancing and receding CLs.
As easily expected, the bulk temperature gets higher with the faster wall speed due to the viscous dissipation as shown in Fig.~\ref{fig:figs4}.

\clearpage

\section{Heat transport around the dynamic CL on the lyophobic wall
\label{sec:AppD}}
The same temperature trend around the dynamic CL shown in Fig.1 with the lyophilic wall in the main text is also 
observed in a system with lyophobic wall.
Specifically, we set a smaller Lennard-Jones energy parameter between solid and fluid molecules 
$\epsilon_{\mathrm{sf}}=0.258\times 10^{-21}\mathrm{J}$ and a faster wall speed $\pm 60\mathrm{m/s}$, 
whereas the other setting is the same as the system in Fig.1.
The static contact angle on such a wall was $\sim 130$\,deg.~\cite{Yamaguchi2019}

\begin{figure}[b]
  \begin{center}
    \includegraphics[width=0.6\linewidth]{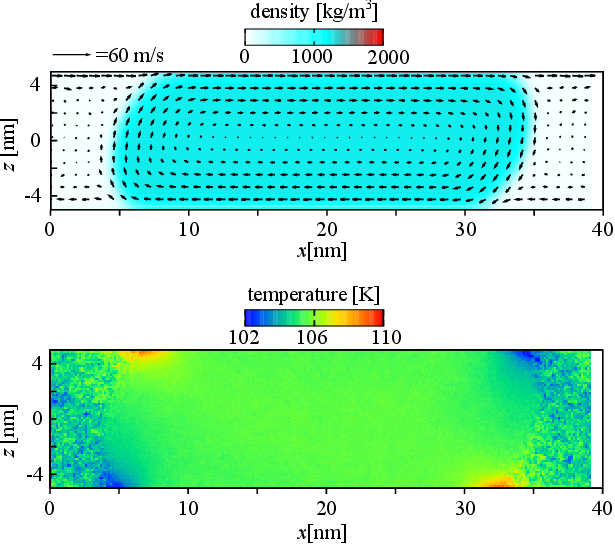}
  \end{center} 
  \caption{\label{fig:figs1}
  (Top) Density distribution and flow field in a system with a lyophobic wall.
  (Bottom) temperature distribution. 
}
\end{figure}

Fig.~\ref{fig:figs1} shows 
the density, velocity and temperature distributions.
The curvature of the menisci is positive unlike the system in Fig.1 due to the low wettability, 
and this results in a positive pressure in the liquid phase.
The bulk liquid temperature is about 106~K which is much higher than the control temperature 85~K due to the high shear rate in bulk liquid, 
whereas that around the advancing CL is about 3~K higher and that around the receding CL is about 4~K lower.
This indicates that the cooling due to the internal energy change around the dynamic CL is induced 
even on the lyophobic wall, and it is more remarkable than that in Fig.2 due to the faster flow.

\clearpage
%\nocite{*}
\bibliography{reference}% Produces the bibliography via BibTeX.